\documentstyle[11pt,moriond,epsfig]{article}

\def\be{\begin{equation}}
\def\ee{\end{equation}}
\def\bea{\begin{eqnarray}}
\def\eea{\end{eqnarray}}

\begin{document}
\title{CHARGE RELAXATION IN THE PRESENCE OF SHOT NOISE IN COULOMB
COUPLED MESOSCOPIC SYSTEMS}

\author{MARKUS B\"UTTIKER}

\address{ D\'epartement de Physique Th\'eorique, Universit\'e de Gen\`eve,\\
CH-1211 Gen\`eve 4, Switzerland}

\maketitle\abstracts{In the presence of shot noise the charge on
a mesoscopic conductor fluctuates. We are interested in the charge 
fluctuations which arise if the conductor is in the proximity of 
a gate to which it is coupled by long range Coulomb forces only. 
Specifically we consider a gate coupled to the edge of a Hall 
bar subject to a quantizing magnetic field 
which contains a quantum point contact. 
The gate is located away from the quantum point contact.
We evaluate the charge relaxation resistance for this geometry.
The charge relaxation resistance determines the current fluctuations 
and potential fluctuations induced into the gate. If there is only one edge 
channel the charge relaxation resistance is determined by transmission and 
reflection probabilities alone, but in the presence of many 
channels the density of states of all edge states 
determines this resistance. This work is to appear in 
"Quantum Physics at Mesoscopic Scale" edited by
D.C. Glattli, M. Sanquer and J. Tran Thanh Van
Editions "Frontieres", 1999.}

\section{Introduction}

Coulomb coupled mesoscopic systems are of interest from a conceptual 
point of view but increasingly also because one of the components 
can serve as a measurement probe testing the other components.
The simplest Coulomb coupled system consists of two conductors,
which form the plates of a mesoscopic capacitor. 
The author, in collaboration with Thomas and Pr\^etre~\cite{btp}
investigated a capacitor which is formed 
by two conductors each 
coupled via a single lead to an electron 
reservoir. In this work it is shown that instead of a simple 
geometrical capacitance the admittance is determined 
by an electrochemical capacitance $C_{\mu}$ and 
a charge relaxation resistance $R_{q}$ which together determine 
the RC-time of the capacitor. The charge relaxation resistance is not simply
a series resistance but instead of transmission probabilities
depends on the energy derivatives of the scattering amplitudes. 
 
If one or both of the conductors are connected to more than one 
reservoir several novel aspects arise. 
Even if we consider only the equilibrium admittance 
the distribution of the displacement currents on the different contacts 
is now an interesting problem which requires a self-consistent 
treatment of the long range Coulomb interactions. 
A dramatic aspect of such a system is the fact that capacitance 
coefficients need no longer to be even functions of the magnetic 
field~\cite{mb}. 
An experiment highlighting 
such asymmetric capacitance coefficients was realized by 
Chen et al.~\cite{chen1} 
who investigated the low frequency admittance of a Hall bar with a gate 
overlapping 
an edge of the conductor. 
Similar experiments in the 
regime of the fractional quantized Hall effect have subsequently also 
been carried out~\cite{chen2,johnson}. We mention here only 
for completeness that at large frequencies an 
arrangement of three or more purely capacitive coupled conductors 
exhibits a dynamic capacitance matrix with elements which are not even functions 
of magnetic field~\cite{christen1}. 
Still a different type of experiment, pioneered by Field et al.~\cite{field}
uses a single electron transistor 
to measure capacitively the electron charge of a mesoscopic conductor 
in proximity to the transistor.

The conductor can be brought into a transport state, if it is 
connected to several leads. 
Instead of equilibrium noise, now the shot noise~\cite{khlus,bu90} due to the 
granularity of the charge, is the source 
of fluctuations. Now the fluctuations of the charge and their relaxation 
are governed by a non-equilibrium charge relaxation 
resistance $R_{v}$. 
Non-equilibrium charge relaxation resistances for fluctuations 
generated by shot noise have been discussed by Pedersen, van Langen 
and the author~\cite{plb}
for quantum point contacts (QPC) and chaotic cavities.
An experiment in which the fluctuations of the charge in the presence 
of shot noise is of paramount importance has been carried out by  
Buks et al.~\cite{buks1}. In this work an Aharonov-Bohm 
ring with a quantum dot which is in turn capacitively coupled to a 
QPC is investigated. With increasing current through 
the QPC, Buks et al. have observed that 
the visibility of the Aharonov-Bohm effect is diminished. 
The charge fluctuations associated with the shot noise of the system 
provide an additional dephasing mechanism.  
In addition to the discussion of the dephasing time given by 
Buks et al.~\cite{buks1} theoretical work by Levinson~\cite{levinson}
and Aleiner~\cite{aleiner1} has addressed this 
problem. The author and A. M. Martin, have related the dephasing time 
in Coulomb coupled conductors to the nonequilibrium 
charge relaxation resistance~\cite{mbam}.   
Here we analyze a simpler geometry: 
We consider a Hall bar with a QPC  which acts as a source of shot noise and consider a capacitively
coupled gate which overlaps the edge of the Hall bar at some distance 
away from the QPC (see Fig. 1).  
This geometry is thus similar to the 
one used in the experiment of Chen et al.~\cite{chen1}. It is also a geometry, which is of 
interest in a new experiment by Buks et al.~\cite{buks2} where instead of a
gate it is planed to couple to a single or double quantum dot. The purpose 
of this work is to present a discussion of the non-equilibrium charge 
relaxation resistance which determines the potential fluctuations
between the conductor and the gate and determines the current induced 
into the gate. The dephasing rate which is of interest in the experiment of 
Buks et al. is the subject of Ref.~\cite{mbam} and will not be discussed
here. 
\begin{figure}
\vspace*{0.5cm}
\epsfxsize=7cm
\centerline{\epsffile{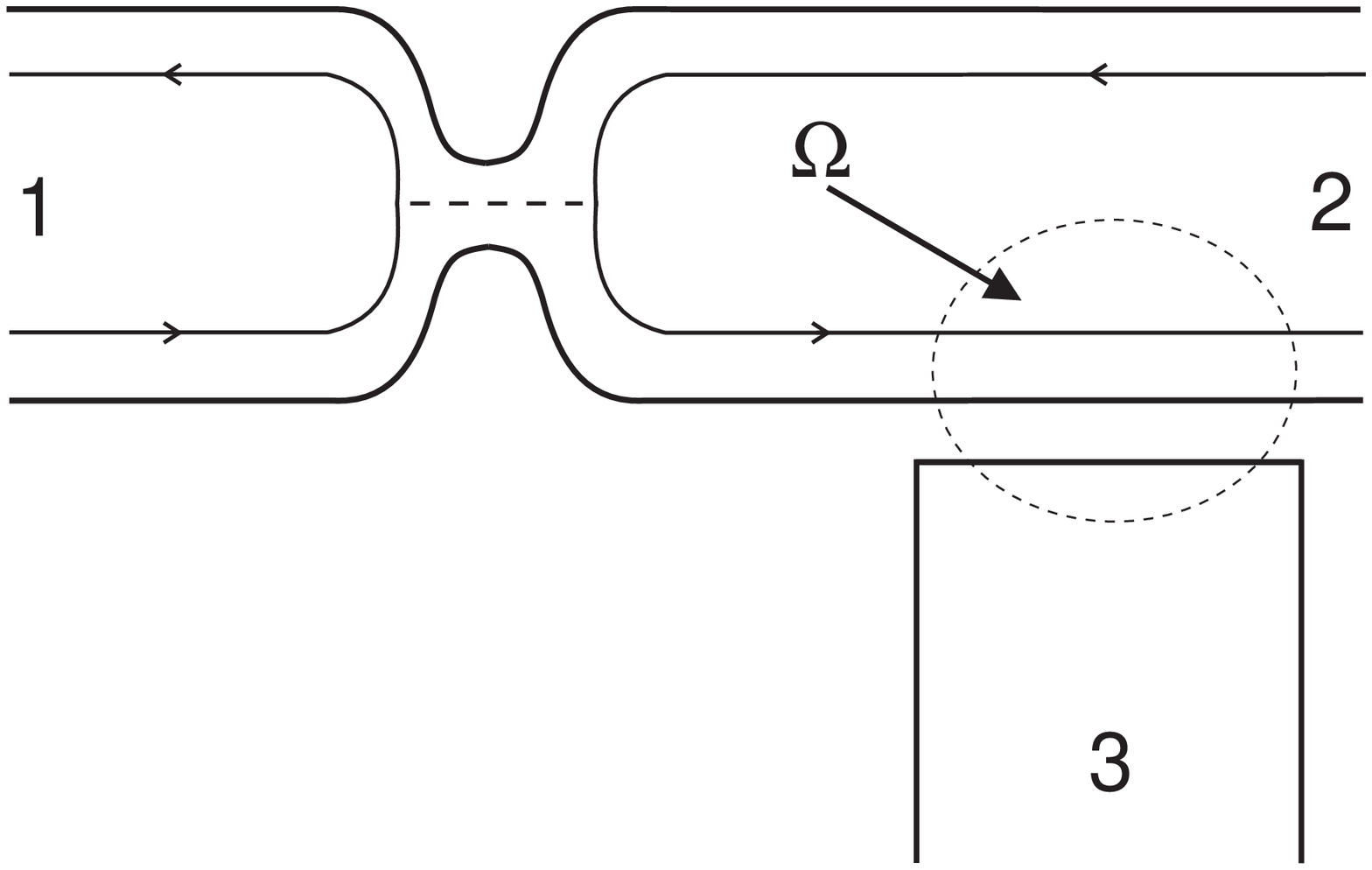}}
\vspace*{0.5cm}
\caption{ \label{qpc_geometry}
Hall bar with a quantum point contact and a gate 
overlapping the edge of the conductor.
}
\end{figure}
\noindent 
\section{The scattering matrix}

To be specific we consider the conductor shown 
in Fig. 1. Of interest is the current
$dI_{\alpha}(\omega)$ at contact $\alpha$ of this conductor 
if an oscillating voltage
$dV_{\beta}(\omega)$ is applied 
at contact $\beta$. Here ${\alpha}$ and ${\beta}$ label the contacts 
of the conductor and the gate and take the values $1,2,3$. Furthermore, 
we are interested in the current noise spectrum 
$S_{I_{\alpha}I_{\beta}}(\omega)$ defined as 
$2\pi S_{I_{\alpha}I_{\beta}}(\omega)\delta(\omega + \omega^{\prime}) = 1/2
\langle {\hat I}_{\alpha}(\omega){\hat I}_{\beta}(\omega^\prime) +
{\hat I}_{\beta}(\omega^\prime){\hat I}_{\alpha}(\omega) \rangle$
and the fluctuation spectrum of the electrostatic potential. 
We assume that 
the charge dynamics is relevant only 
in the region underneath the gate. Everywhere else we 
assume the charge to be screened completely. This is a strong assumption: 
The QPC is made with the help
of gates (capacitors) and also exhibits its own capacitance~\cite{christen2}. 
Edge states might generate long range fields, etc. Thus the results presented 
below can only be expected to capture the main effects but can certainly 
be refined. We assume that the gate is a macroscopic 
conductor and
screens perfectly. 
To be brief, we consider the case of one edge channel only. 

The scattering matrix of the QPC alone can be described by 
$r \equiv s_{11} = s_{22} = - i {\cal R}^{1/2}$
and $t \equiv s_{21} = s_{12} = {\cal T}^{1/2}$
where ${\cal T} = 1- {\cal R}$ is the transmission probability 
through the QPC. 
Here the indices $1$ and $2$ label the reservoirs of Fig. 1. 
A carrier traversing the region underneath the gate 
acquires a phase $\phi(U)$ which depends on the 
electrostatic potential $U$ in this region. 
Since we consider only the charge pile up in this region 
all additional phases
in the scattering problem are here without relevance. The total scattering
matrix of the QPC and the traversal of the region $\Omega$ is then simply
\begin{equation} \label{sm}
{\bf s} = 
\left( \matrix{ r & t \cr 
t e^{i\phi} & r e^{i\phi} } \right).
\end{equation} 
If the polarity of the magnetic field is reversed 
the scattering matrix is given by $s_{\alpha\beta}(B) = s_{\beta\alpha}(-B) $, 
i. e. in the reversed magnetic field it is only the second column
of the scattering matrix which contains the phase $\phi(U)$. 
In what follows the dependence of the scattering matrix on the phase $\phi$
is crucial. We emphasize that the approach presented here can be generalized 
by considering all the phases of the problem and by 
considering these phases and the amplitudes to depend 
on the entire electrostatic potential landscape.

\section{Density of States Matrix Elements}

To describe the charge distribution due to carriers 
in an energy interval $dE$ in our conductor, 
we consider a current of unit amplitude incident from contact 
$\beta$ and simultaneously a current of unit amplitude 
incident from contact $\gamma$ which leave the conductor 
through contact $\alpha$. This gives rise to a charge 
in the region of interest given by $N_{\beta\gamma} (\alpha) dE$ 
where $dE$ is a small energy interval, and $N_{\beta\gamma} (\alpha)$
is a density of states matrix element given by 
\begin{equation} \label{elmdef}
N_{\beta \gamma} (\alpha) = 
- (1/4\pi i) 
[s^{\ast}_{\alpha\beta} (\partial s_{\alpha \gamma}/\partial eU) -
(\partial s^{\ast}_{\alpha \beta}/\partial eU) s_{\alpha \gamma} ] .
\end{equation} 
The low frequency charge dynamics can be found if these
density of states matrix elements are known. 
For the specific example given by Eq. (\ref{sm}) 
we find that all elements with $\alpha =1$ vanish:
$N_{11} (1) =  
N_{21} (1) = N_{12} (1) = N_{22} (1) = 0$. 
There are no carriers incident from contact $1$ or $2$ 
which pass through region $\Omega$ and leave the conductor 
through contact $1$. The situation is different if we demand 
that the current leaves the sample through contact $2$. 
Now we find 
\begin{equation} \label{Nm}
N_{\beta\gamma}(2)  = 
\left( \matrix{ {\cal T}N & t^{\ast} r N \cr 
r^{\ast} t N & {\cal R}N} \right),
\end{equation} 
where $N = - (1/2e\pi) d\phi/dU $ is the density of states 
of carriers in the edge state underneath the gate. 
For the reverse magnetic field polarity
all components of the matrix vanish except 
the elements $N_{22} (1) = {\cal T}N$ and $N_{22} (2) = {\cal R}N$. 

For the charge and its fluctuations underneath 
the gate it is not relevant through which contact carriers leave. 
The charge pile up and its fluctuations are thus governed by a matrix 
\begin{equation} \label{qdenm}
N_{\beta \gamma} = \sum_{\alpha} N_{\beta\gamma} (\alpha)
\end{equation} 
which is obtained by summing over the contact index $\alpha$ 
from the elements given by Eq. (\ref{elmdef}). 
For our example the density matrix elements for the charge are thus 
evidently given by $N_{\beta \gamma} = N_{\beta\gamma}(2)$ 
whereas for the reversed magnetic field polarity 
we have $N_{11} = {\cal T} N$, $N_{22} = {\cal R}N$ and $N_{21} = N_{21} = 0$. 

Furthermore, we will make use of the {\em injectivity} of a contact 
into the region $\Omega$ and will make use of the emissivity of
the region $\Omega$ into a contact. 
The injectivity of contact is the charge injected into a region 
in response to a voltage variation at the contact $\alpha$,
independently through which contact the carriers leave the sample. 
The injectivities of contact $1$ and $2$ are 
\begin{equation}
N_{1} =  N_{11}(1) + N_{11} (2) = {\cal T} N
\label{in1}
\end{equation}
\begin{equation}
N_{2} =  N_{22}(1) + N_{22} (2) = {\cal R} N
\label{in2}
\end{equation}
Note that the sum of the injectivities of both contacts is just the density
of states $N$ underneath the gate. 
The {\em emissivity} of region $\Omega$ is the 
portion of the density of states of carriers 
which leave the conductor through contact $\alpha$ 
irrespectively from which contact they entered
the conductor. We find emissivities
\begin{equation}
N^{(1)} =  N_{11}(1) + N_{22} (1) = 0,  
\label{e1}
\end{equation}
\begin{equation}
N^{({2})} =  N_{11}(2) + N_{22} (2) = N . 
\label{e2}
\end{equation} 
Any charge accumulation or depletion is only felt in contact $2$.
The injectivities and emissivities in the magnetic field 
$B$ are related by reciprocity to the 
emissivities and injectivities in the reversed magnetic field, 
$N_{\alpha}(B) = N^{(\alpha)}(-B)$ and 
$N^{(\alpha)}(B) = N_{\alpha}(-B)$.
In contrast, the density of states
$N$ is an even function of magnetic field. 

\section{The Poisson Equation: The effective interaction}

Thus far we have only considered bare charges. The true charge, 
however, is determined by the long range Coulomb interaction. 
First we consider the screening of the average charges and in a second 
step we consider the screening of charge fluctuations. We describe the 
long range Coulomb interaction between the charge on the edge state 
and on the gate with the help of a geometrical capacitance $C$. 
The charge on the edge state beneath the gate is determined by the voltage 
difference between the edge state and the gate $dQ = C(dU - dV_{g})$,
where $dU$ and $dV_{g}$ are deviations 
from an equilibrium reference state.
On the other hand the charge beneath the gate can also be expressed 
in terms of the injected charges  $e^{2} N_{1} dV_{1}$  in response 
to a voltage variation at contact $1$ and $e^{2} N_{2} dV_{2}$
in response to a voltage variation at contact $2$. 
The injected charge leads to a response in the internal potential 
$dU$ which in turn generates a screening charge $-e^{2}NdU$
proportional to the density of states. 
Thus the Poisson equation for the charge underneath the gate is 
\begin{equation}\label{poisson1}
dQ = C (dU - dV_{g}) = e^{2} N_{1} dV_{1} + e^{2} N_{2} dV_{2} 
- e^{2}N dU
\end{equation}
and the charge on the gate is given by $- dQ = C (dV_{g} - dU)$. 
Solving Eq. (\ref{poisson1}) for $dU$ gives 
\begin{equation}\label{u1}
dU = G_{eff} (C dV_{g} + e^{2} N_{1} dV_{1} + e^{2} N_{2} dV_{2}), 
\end{equation}
where $G_{eff} = (C +e^{2}N)^{-1}$ is an effective interaction 
which gives the potential underneath the gate in response to an increment 
in the charge. 

\section{Admittance}

Consider now the low-frequency conductance:
To leading order in the frequency $\omega$ we write 
\begin{equation} \label{admit}
G_{\alpha\beta}(\omega) = G_{\alpha\beta} (0) - i \omega E_{\alpha\beta}
+ \omega^{2}  K_{\alpha\beta} + O(\omega^{3}) .
\end{equation}
Here  $G_{\alpha\beta} (0)$ is the dc-conductance matrix,  
$E_{\alpha\beta}$ is the {\em emittance} matrix, and $K_{\alpha\beta}$
is a second order dissipative contribution to the frequency dependent 
admittance. 
The zero-frequency dc-conductance matrix has only four non-vanishing
elements which are given by 
$G \equiv G_{11} = G_{22} = - G_{12} = - G_{21} = (e^{2}/h) {\cal T}$. 
Ref.~\cite{mb} showed that the emittance matrix ${\bf E}$ is given by 
\begin{equation} \label{Em}
E_{\alpha\beta} = e^{2} N_{\beta\beta}(\alpha) - e^{2} N^{(\alpha)}G_{eff}N_{\beta}
\end{equation}
As it is written, Eq. (\ref{Em}) applies only to the elements 
where $\alpha$ and $\beta$ take the values $1$ or $2$. 
The remaining elements can be obtained from current conservation 
(which demands that the elements of each row 
and column of this matrix add up to zero) or can be obtained directly
by using a more general formula. For our example we find an emittance matrix, 
\begin{equation} \label{em}
{\bf E} = C_{\mu}
\left( \matrix{  0 &   0 &  0 \cr 
                 {\cal T} &   {\cal R} & -1 \cr  
               - {\cal T} & - {\cal R} &  1} \right),
\end{equation}
with an electrochemical capacitance of the conductor vis-a-vis the gate 
given by $C_{\mu} = C e^{2}N/(C+e^{2}N)$. 
Eq. (\ref{em}) determines the displacement currents 
in response to an oscillating voltage at one of the contacts. 
There is no displacement current at contact $1$
(the elements of the first row vanish) which is consequence of our 
assumption that charge pile up occurs only underneath the gate. 
The emittance matrix in the magnetic field $B$ and in the magnetic 
field $-B$ are related by reciprocity, 
$E_{\alpha\beta}(B) = E_{\beta\alpha}(-B)$. 
For the reverse polarity, a voltage oscillation at contact $1$ 
generates no displacement currents (the elements of the first column 
vanish). 

The emittance matrix element $E_{21}$ is positive and thus 
has the sign not of a capacitive but of an inductive 
response. The elements of row $3$ and column $3$ are a consequence of 
purely capacitive coupling and have the sign associated with the 
elements of a capacitance matrix. Thus these elements 
represent the capacitance matrix elements which can be measured 
in an ac-experiment. Note that the capacitances $E_{31} \equiv C_{31}$
and $E_{32} \equiv C_{32}$ depend not only on the density of states 
and geometrical capacitances but also on transmission and reflection
probabilities. Measurement of these capacitances provides
thus a direct confirmation of the concept of {\em partial} density 
of states~\cite{mb,christen2}. Furthermore, we see that for instance $C_{31} (B)  \equiv 
E_{31} = {\cal T} C_{\mu}$ but $C_{31} (-B) = 0$. A similarly striking variation 
of the capacitance coefficients was observed in the experiment of Chen et
al.~\cite{chen1} in the integer quantum Hall effect and in 
Refs.~\cite{chen2,johnson} in the fractional quantum Hall effect. 

\section{Bare charge fluctuations}

Let us now turn to the charge fluctuations. With the help of 
the charge density matrix the low frequency limit of the 
bare charge fluctuations can be obtained~\cite{btp,plb,mbam}. It is given by 
\begin{equation}
    S_{NN}(\omega) = h 
        \sum_{\delta\gamma} \int dE\, F_{\gamma\delta}(E,\omega) 
        N_{\gamma\delta}(E,E+\hbar\omega)
        N^{\dagger}_{\gamma\delta} (E,E+\hbar\omega) 
\label{qfluct1}
\end{equation}
where the elements of $N_{\gamma\delta}$ are in the zero-frequency limit 
of interest here given by Eq. (\ref{qdenm}) and  
and $F_{\gamma\delta} = f_{\gamma}(E)(1-f_{\delta}(E+\hbar \omega))
+ f_{\delta}(E+\hbar \omega)(1- f_{\gamma}(E))$
is a combination of Fermi functions. 
Using only the zero-frequency 
limit of the elements of the charge operator determined above gives, 
\begin{eqnarray}\label{qfluct2}
    S_{NN}(\omega) =  h N^{2} &[ & {\cal T}^{2} \int dE\ F_{11}(E,\omega) 
                            + {\cal T}{\cal R} \int dE\ F_{12}(E,\omega)\nonumber\\
                         &+ & {\cal T}{\cal R} \int dE\ F_{21}(E,\omega)
                            + {\cal R}^{2} \int dE\ F_{22}(E,\omega)] .
\end{eqnarray}
At equilibrium all the Fermi functions are identical and we obtain 
$ S_{NN}(\omega) = h N^{2} \int dE\ F(E,\omega)$ 
which in the zero-frequency limit is 
\begin{eqnarray}
    S_{NN}(\omega) = h N^{2} kT 
\label{qfluct3a}
\end{eqnarray}
and at zero-temperature to leading order in frequency is, 
\begin{eqnarray}
    S_{NN}(\omega) = h N^{2} \hbar \omega .
\label{qfluct3b}
\end{eqnarray}
In the zero-temperature, zero-frequency limit,
in the presence of a current through the sample, we find 
for the charge fluctuations associated with shot noise 
\begin{eqnarray}
    S_{NN}(\omega) = h N^{2} {\cal T}{\cal R}  e|V| .
\label{qfluct4}
\end{eqnarray}
However, the bare charge fluctuations are not by themselves 
physically relevant.

\section{Fluctuations of the true charge}

To find the fluctuations of the true charge 
we now write the Poisson equation for the fluctuating charges. 
All contact potentials are at their equilibrium value, 
$dV_{1} = dV_{2} = dV_{g} = 0$. 
The fluctuations of the bare charge now generate fluctuations
in the electrostatic potential. Thus the electrostatic potential 
has also to be represented by an operator ${\hat U}$. Furthermore, the potential 
fluctuations are also screened. As in the case of the average charges we take 
the screening to be proportional to the density of states $N$ but replace 
the c-number $U$ by its operator expression ${\hat U}$. 
The equation for the fluctuations of the true charge is thus 
\begin{equation} \label{poisson2}
d {\hat Q} = C d {\hat U}  = e\hat{\cal N} - e^{2}N {\hat U}
\end{equation}
whereas the fluctuation of the charge on the gate is simply
$ - d \hat Q = - C d \hat U$. 
Thus $d\hat Q$ is the charge operator of the {\em dipole} which forms 
between the charge on the edge state and the charge on the gate. 
Solving Eq. (\ref{poisson2})
for the potential operator ${\hat U}$ and using this result 
to find the fluctuations of the charge $d{\hat Q}$ gives 
\begin{equation}\label{qfluct1b}
    S_{QQ}(\omega) = e^{2} C^{2}G^{2}_{eff}  S_{NN}(\omega) 
                   =  2C^{2}_{\mu} (1/2e^{2}) 
                   (S_{NN}(\omega)/N^{2}) .
\end{equation}
We now discuss three limits of this result. 

\section{Equilibrium and non-equilibrium charge relaxation resistance} 

At equilibrium, in the zero-frequency limit, the charge fluctuation 
spectrum can be written with the help of the equilibrium charge relaxation
resistance $R_{q}$,  
\begin{eqnarray}
    S_{QQ}(\omega) = 2C^{2}_{\mu} R_q kT.  
\label{qflucteq}
\end{eqnarray}
For our specific example we find using Eqs. (\ref{qfluct3a}) and 
(\ref{qfluct1b}),
\begin{equation} \label{rq}
R_q = h/2e^{2} .
\end{equation} 
The charge relaxation resistance is universal 
and equal to {\em half} a resistance quantum as expected for a single
edge state~\cite{christen3}. 
At equilibrium the fluctuation spectrum is via the fluctuation 
dissipation theorem directly related to the dissipative part 
of the admittance. We could also have directly evaluated 
the element $K_{33}$ of Eq. (\ref{admit}) 
to find $K_{33} = C_{\mu}^{2} R_{q}$.
Second at equilibrium, but for frequencies which are large 
compared to the thermal energy, but small compared to any intrinsic excitation
frequencies, we find that zero-point fluctuations give rise 
to a noise power spectral density
\begin{eqnarray}
    S_{QQ}(\omega) = 2C^{2}_{\mu} R_q \hbar \omega  
\label{qfluctz}
\end{eqnarray}
which is determined by the charge relaxation resistance Eq. (\ref{rq}). 
Third, in the presence of transport, we find in the zero-frequency,
zero-temperature limit, 
a charge fluctuation spectrum determined by the non-equilibrium charge relaxation
resistance $R_v$, 
\begin{eqnarray}
    S_{QQ}(\omega) = 2C^{2}_{\mu} R_{v} e|V|, 
\label{qfluctv}
\end{eqnarray}
where $|V|$ is the voltage applied between the two contacts of the sample
and a non-equilibrium charge relaxation resistance 
\begin{eqnarray}
      R_{v} = (h/e^{2}) {\cal T}{\cal R}, 
\label{rqv}
\end{eqnarray}
which is maximal for a semi-transparent QPC.  
We emphasize that the universality of the equilibrium charge relaxation
resistance and the property that the non-equilibrium charge relaxation
resistance follows directly the low frequency shot noise spectrum 
are due to the special geometry investigated here and are due 
to the fact that we considered the case of one edge state only. 
If two or more edge states are present both $R_q$ and $R_v$ depend 
on the density of states of all the edge channels underneath the 
gate~\cite{mbam}. 

The current at the gate due to the charge fluctuations 
is $dI_{g} = -i\omega dQ(\omega)$ and thus its fluctuation 
spectrum is given by  
$S_{I_{g}I_{g}}(\omega) = \omega^{2} S_{QQ}$. The potential 
fluctuations are related to 
the charge fluctuations by $d{\hat U} =d{\hat Q}/C$ and thus 
the spectral density of the potential 
fluctuations is $S_{UU}(\omega) = C^{-2} S_{QQ}$. 
Thus the charge relaxation resistance determines, together with 
the electrochemical and geometrical capacitance, 
the fluctuations of the charge,
the potential and the current induced into the gate. Since dephasing 
rates can be linked to the low frequency limit of the potential 
fluctuations~\cite{levinson} the charge relaxation 
resistance also determines the 
dephasing rate in Coulomb coupled mesoscopic conductors~\cite{mbam}.

\section{Acknowledgments}

This work was supported by the Swiss National Science Foundation
and by the TMR network Dynamics of Nanostructures.\\

\end{document}